\documentclass[twocolumn,showpacs,reprint,aps,prl]{revtex4-1}
\usepackage{graphicx}
\usepackage{subfigure}
\usepackage{rotating}
\usepackage{amsmath, amssymb, mathtools}
\usepackage{xspace}
\usepackage{titlesec}
	\titleformat{\section}[runin]{\normalfont\normalsize\bfseries}{\thesection}{0pt}{}[.]
	\titlespacing{\section}{0pt}{\parskip+0.5ex}{2ex}

\newcommand{\tf}{t_{\mathrm{f}}}

\newcommand{\sym}{\operatorname{Sym}}
\newcommand{\st}{\varepsilon}

\newcommand*{\cf}{\mbox{c.\,f.}\xspace}

\newcommand*{\ie}{\mbox{i.\,e.}\xspace}
\newcommand*{\resp}{\mbox{resp.}\xspace}

\begin{document}


\title{How nanomechanical systems can minimize dissipation}
\date{\today}

\author{Paolo Muratore-Ginanneschi}
\email{paolo.muratore-ginanneschi@helsinki.fi}
\author{Kay Schwieger}
\email{kay.schwieger@helsinki.fi}
\affiliation{University of Helsinki, Department of Mathematics and Statistics
    P.O. Box 68 FIN-00014, Helsinki, Finland}

\begin{abstract} Information processing machines at the nanoscales are
unavoidably  affected  by   thermal  fluctuations.   Efficient  design
requires understanding how nanomachines  can operate at minimal energy
dissipation.  In this letter we focus on mechanical systems controlled
by smoothly  varying potential forces.   We show that  optimal control
equations way if the energy cost  to manipulate the potential is taken
into  account.  When such  cost  becomes  negligible, optimal  control
strategy  can be  constructed by  transparent geometrical  methods and
recovers  the solution  of  optimal mass  transport  equations in  the
overdamped  limit.  Our  equations  are equivalent  to hierarchies  of
kinetic  equations  of a  form  well-known  in  the theory  of  dilute
gases.  From  our results,  optimal  strategies  for energy  efficient
nanosystems  may be  devised  by established  techniques from  kinetic
theory.
\end{abstract}

\pacs{05.40.-a, 05.70.Ln, 02.30.Yy, 05.20.Dd, 02.50.Ey}

\keywords{Brownian   motion,  free   energy,   protocols,  statistical
  mechanics,   stochastic   processes,   stochastic  control   theory,
  thermodynamics}
\maketitle


Recent experiments with colloidal  particles, and integrated platforms
of nanomagnetic memory and logic circuits exhibited the possibility to
design  and control  information  processing machines  on a  molecular
scale  \cite{ToSaUeMuSa10,LaCaBo11,BeArPeCiDiLu12}. These  experiments
are  a  first  experimental  step  towards  low  dissipation  Brownian
computers, a concept theoretically envisaged decades ago \cite{Ben82}.
Furthermore,  these  experiments  put  to  test  new  developments  of
non-equilibrium thermodynamics  such as refinements of  the Second Law
stemming         from         fluctuation        theorems         (see
e.g.    \cite{EvSe94,GaCo95,Jar97,Cro97,JiangQianQian}     and    also
\cite{DiLu09,AuGaMeMoMG12}).    Engineering   nanomachines,   however,
remains technologically challenging \cite{VdBr10} and demands a better
theoretical understanding of how energy dissipation can be minimized.

The   Langevin--Kramers  dynamics   is   the   reference  model   (\cf
\cite{vanKampen})   epitomizing  effects   in  nanosystem   mechanics:
kinetic-plus-potential  Hamiltonian, mechanical  friction by  a Stokes
drag force, and thermal noise. In the Langevin--Kramers framework, the
Second Law of thermodynamics is amenable to a mathematical formulation
in  terms   of  an  adapted  \emph{Schr\"odinger   diffusion  problem}
\cite{Aebi}: given  the phase  space probability  densities describing
the state  of the system at  finite initial and final  times, find the
potential force which steers the  initial into the final density while
minimizing the average energy dissipation.

In  this letter  we conceptualize  the Schr\"odinger  diffusion as  an
optimal stochastic control problem  \cite{FlemingSoner}. Our aim is to
derive the  optimal control strategy  over the class  of \emph{smooth}
potential  forces  for  the  problem  adapted to  the  Second  Law  of
Thermodynamics.  From the experimental  slant, smooth potentials model
macroscopic degrees of freedom of the system whose state is determined
by  external sources  \cite{AlRiRi11}.  We  show that  our problem  is
well-posed  when regarded  as  the  limit of  a  more general  control
problem which takes into account the  energy cost of the control.  The
phase-space optimal control  equations turn out to be  amenable to the
form of \emph{Bogoljubov--Born--Green--Kirkwood--Yvon kinetic momentum
hierarchies}, which are  well-known in the theory of  dilute gases (see
e.g. \cite{CeIlPu94}). This is an  important observation as it renders
immediately  available  a  toolbox  of mathematical  methods  for  the
analysis of our optimal  control equations \cite{Lev96}.  In agreement
with  physical intuition,  in  the overdamped  regime  we recover  the
Monge--Amp\`ere--Kantorovich   equations~\cite{Villani},  which   were
recently  proved  to govern  minimal  dissipation  transitions in  the
Langevin--Smoluchowski modeling \cite{AuMeMG11}.  Moreover, elementary
arguments    show   that    the    solution   of    the   very    same
Monge--Amp\`ere--Kantorovich equations yield a general lower bound for
the average energy dissipation by smooth Langevin--Kramers dynamics.
 
Finally, we notice  that our results are  based on \mbox{Pontryagin}'s
principle (see  \cite{BoPi05} for a  concise review) in  a formulation
inspired  by  \cite{GuMo83}.   To  the  best  of  our  knowledge,  our
formulation  is  a novelty  which  may  be  of relevance  for  control
problems in other disciplines.

\section{Model}

We  define the  kinetic-plus-potential  Langevin--Kramers dynamics  by
means of its scalar generator
\begin{equation*}
\mathfrak{L}=\frac{\boldsymbol{p}}{m}\cdot\partial_{\boldsymbol{q}}-\left(\frac{\boldsymbol{p}}{\tau}
+\partial_{\boldsymbol{q}}U\right)\cdot\partial_{\boldsymbol{p}}+\frac{m}{\beta\,\tau}\partial_{\boldsymbol{p}}^{2}
\end{equation*} Here $\beta$ is the inverse of the temperature, $m$ is
the mass of a Brownian particle under a time dependent potential force
$\partial_{\boldsymbol{q}}U\equiv(\partial_{\boldsymbol{q}}U)(\boldsymbol{q},t)$,
and $\tau$ is  the characteristic time of the Stokes  drag. We suppose
the dynamics  to occur  on an  $2d$-dimensional Euclidean  phase space
with                 coordinates                 $\boldsymbol{x}\equiv
\left[\boldsymbol{q}\,,\boldsymbol{p}\right]$,  where $\boldsymbol{q}$
and $\boldsymbol{p}$  denote as usual  positions and momenta.   By the
generator  $\rho\equiv\rho\left(\boldsymbol{x},t\right)$ evolves  then
according to the Fokker--Planck equation
\begin{eqnarray}
\label{FP} (\partial_{t}-\mathfrak{L}^{\dagger})\rho=0
\end{eqnarray}      where     $\mathfrak{L}^{\dagger}$      is     the
$\mathbb{L}^{2}(\mathbb{R}^{2d})$   adjoint  of   $\mathfrak{L}$  with
respect   to   the   Lebesgue   measure.   Stochastic   thermodynamics
considerations  (see  e.g.   \cite{JiangQianQian,ChGa08})  uphold  the
interpretation of
\begin{equation*}
\mathcal{Q}=-\frac{d\,\tf}{\beta\,\tau}+\int_{0}^{\tf}\mathrm{d}t\int_{\mathbb{R}^{2d}}
\mathrm{d}^{2d}x\,\rho\frac{\|\boldsymbol{p}\|^{2}}{m\,\tau}
\end{equation*}  
as the  mean heat  release by  the Brownian  particle
during the time interval $[0,\tf]$. The  mean heat release is given by
the expected value of the line integral over the Stokes drag
\begin{eqnarray}
\label{dissipation}
\mathcal{E}=\int_{0}^{\tf}\mathrm{d}t\int_{\mathbb{R}^{2
d}}\mathrm{d}^{2 d}x\,\rho\frac{\|\boldsymbol{p}\|^{2}}{m\,\tau}
\end{eqnarray} 
minus its  Maxwell--Boltzmann equilibrium equipartition
value.   Hence, finding  a tight  lower bound  for the  Second Law  of
thermodynamics over finite-time transitions governed by (\ref{FP}) and
transforming     $\rho(\boldsymbol{x},0)=\rho_{\iota}(\boldsymbol{x})$
into  $\rho(\boldsymbol{x},\tf)=\rho_{\mathrm{f}}(\boldsymbol{x})$  is
equivalent  to   finding  a  Schr\"odinger  diffusion   process  which
minimizes (\ref{dissipation})  by an optimal  choice of $U$.   In this
letter, we  will always  consider boundary conditions  compatible with
thermal       equilibrium:        $\rho_{\mathrm{j}}(\boldsymbol{x})       =
\mu_{\mathrm{j}}(\boldsymbol{q})
\mu_{\scriptscriptstyle{MB}}(\boldsymbol{p})      $,     $\mathrm{j}      =
\mathrm{\iota},\mathrm{f}$                                        with
$\mu_{\scriptscriptstyle{MB}}(\boldsymbol{p})=[\beta/(2\,\pi\,
m)]^{d/2}
\exp\left\{-\frac{\beta\,\|\boldsymbol{p}\|^{2}}{2\,m}\right\}$    the
Maxwell--Boltzmann momentum distribution.

\section{Bounds} 

Elementary statistical moment inequalities immediately yield
\begin{eqnarray}
\label{bound}
\mathcal{E}\geq\int_{0}^{\tf}\frac{\mathrm{d}t}{\tau}\int_{\mathbb{R}^{d}}
\mathrm{d}^{d}q\,\mu\,m\,\|\boldsymbol{v}\|^{2}
\end{eqnarray}         with        the         marginal        density
$\mu(\boldsymbol{q},t)                                 \equiv
\int_{\mathbb{R}^{d}}\mathrm{d}^{d}p\,\rho(\boldsymbol{q},\boldsymbol{p},t)$
and  the  macroscopic  velocity  (or  first  order  kinetic  cumulant)
$\boldsymbol{v}(\boldsymbol{q},t)\equiv
[\mu(\boldsymbol{q},t)\,m]^{-1}
\int_{\mathbb{R}^{d}}\mathrm{d}^{d}p\,\rho(\boldsymbol{q},\boldsymbol{p},t)\,\boldsymbol{p}$.
A  well-known result  of kinetic  theory \cite{CeIlPu94}  implies that
$\mu$   obeys  a   continuity   equation   with  respect   to
$\boldsymbol{v}$.  Hence we obtain (see e.g. \cite{PMG14} and below) a
lower bound  for the  right hand  side of  (\ref{bound}) if  we choose
$\boldsymbol{v}$  to  be  solution of  a  Monge--Amp\`ere--Kantorovich
system     (Burgers     plus     mass     continuity)     transporting
$\mu_{\iota}$ into  $\mu_{\mathrm{f}}$ in $[0,\tf]$.
Repeating  analogous considerations  on conditional  position averages
yields    for    equilibrium    boundary    conditions    the    bound
$\mathcal{Q}\,\geq\,0$.   In general,  there  is no  reason to  expect
these  simple  bounds   to  be  tight.  Moreover,   the  knowledge  of
$\boldsymbol{v}$ does not specify an optimal control $U$.

\section{Optimal Control}

To tackle the optimal control problem 
we construct from (\ref{dissipation}) the
\emph{energy cost} functional
\begin{equation}
	\label{cost}
	\mathcal{A} = \mathcal{E}
	+\frac{2\,g}{\beta}\mathcal{D}\left(\rho\| \bar{\rho}\right)
	-\int_{0}^{\tf}\hspace{-0.1cm}\mathrm{d}t\int_{\mathbb{R}^{2 d}}\hspace{-0.1cm}
	\mathrm{d}^{2 d}x\,V\left(\partial_{t}-\mathfrak{L}^{\dagger}\right)\rho
\end{equation}
where $g\geq 0$ and, upon denoting the local equilibrium potential by $S(\boldsymbol{q},t)=
-\ln\frac{\tau^{d}\,\mu(\boldsymbol{q},t)}{\beta^{d/2}\,m^{d/2}}$, 
we define
\begin{equation*}
	\mathcal{D}\left(\rho \, \| \, \bar{\rho}\right)
	\equiv
	\frac{\tau\,\beta}{2\,m}\int_{0}^{\tf}\hspace{-0.1cm}\mathrm{d}t\hspace{-0.05cm}\int_{\mathbb{R}^{2 d}}
	\hspace{-0.1cm}\mathrm{d}^{2 d}x\,\rho\,
	\|\partial_{\boldsymbol{q}}(U-\tfrac{k}{\beta}\,S)\|^{2}
\end{equation*}
with a real constant~$k$. 

The   last  term   in  (\ref{cost})   encapsulates  \emph{Pontryagin's
principle}.   Namely,  the  term  vanishes  if  $\rho$  satisfies  the
Fokker-Planck equation  \eqref{FP}. Thus, the dynamics  is enforced by
the  Lagrange  multiplier   $V\equiv  V(\boldsymbol{x},t)$,  which  we
interpret as  the value function  of Bellman's formulation  of optimal
control \cite{FlemingSoner}.  The parameter  $g\,\geq\, 0$ couples the
energy dissipation $\mathcal{E}$ to  a term $\mathcal{D}$ modeling the
energy cost of  the control.  Namely, $\mathcal{D}\left(\rho  \, \| \,
\bar{\rho}\right)$ is  the Kullback-Leibler divergence  between $\rho$
of the  process and  the density  $\bar{\rho}$ of  a Langevin--Kramers
process    driven   by    the   potential    $\smash{\tfrac{k}{\beta}}
\,S$.  As $\smash{\tfrac{1}{\beta}} \, S$ is the local
equilibrium   potential    for   the   position    marginal   density,
$\mathcal{D}\left(\rho  \,   \|  \,  \bar{\rho}\right)$   vanishes  at
equilibrium for $k=1$.  The case $k=0$ describes  instead the relative
entropy   between  the   controlled  and   the  uncontrolled   ($U=0$)
Langevin--Kramers dynamics.
 
Following Pontryagin's  principle, we look  for extremal value  of the
energy    cost    functional    (\ref{cost})   versus    the    triple
$(\rho,V,U)$. The  calculation is straightforward but  relies on three
crucial observations.  First, for  any $g\,>\,0$,  the energy  cost is
convex   in  the   control   $\partial_{\boldsymbol{q}}U$  and   hence
coercive~\cite{FlemingSoner}.  Second,  the  energy cost  depends  non
linearly   upon~$\rho$    via   the   local    equilibrium   potential
$S$. Third, since $U$ depends only upon position variables, we
can  average out  momenta  from energy  cost  variations with  respect
to~$U$.  We thus  arrive to  the  system of  three extremal  equations
formed by:  first, the Fokker-Planck equation~\eqref{FP};  second, the
dynamic programming equation
\begin{eqnarray}
	\label{dp}
	(\partial_{t}+\mathfrak{L})V+\frac{\parallel\boldsymbol{p}\parallel^{2}}{m\,\tau}
	+\frac{2\,g}{\beta}\mathcal{D}_{\rho}^{\prime}=0
\end{eqnarray}
with $\mathcal{D}_{\rho}^{\prime}\equiv
\frac{\beta\,\tau}{2\,m}\,\{\|\partial_{\boldsymbol{q}}(U-\frac{k}{\beta}S)\|^{2}
-\frac{2\,k}{\beta}\partial_{\boldsymbol{q}}^{2}(U-\frac{k}{\beta}S)
\}$; and third, the equation for the control potential
\begin{equation}
	\label{potential}
	\nabla_{\boldsymbol{q}}^{\scriptscriptstyle{S}}\cdot
	\bigl( \tfrac{m}{2\tau} \, \mathsf{V}^{\scriptscriptstyle{(1)}}
	- g\,\partial_{\boldsymbol{q}} (U - \tfrac{k}{\beta} S) \bigr)
	=0
\end{equation}
where we write 
$\nabla_{\boldsymbol{q}}^{\scriptscriptstyle{S}}\equiv\partial_{\boldsymbol{q}}-(\partial_{\boldsymbol{q}}S)$ 
and define
\begin{eqnarray}
	\label{force}
	\mathsf{V}^{\scriptscriptstyle{(1)}}(\boldsymbol{q},t)=
	\int_{\mathbb{R}^{d}}\mathrm{d}^{d}p\,
	\frac{\rho(\boldsymbol{q},\boldsymbol{p},t)}{\mu(\boldsymbol{q},t)}\,
	\partial_{\boldsymbol{p}}V(\boldsymbol{q},\boldsymbol{p},t)
\end{eqnarray}
These three equations, the optimal control system, is the first result of this letter. 
We now turn to analyze its main physical consequences.

\section{Kinetic hierarchies and controllability} 

In general, we can turn the energy cost \eqref{cost} 
into an average restricted to configuration space by introducing conditional 
momentum cumulants of any order $n$:
\begin{equation*}
	\mathsf{F}_{\boldsymbol{i}_{n}}^{\scriptscriptstyle{(n)}}(\boldsymbol{q},t) 
	\equiv
	\left.\frac{\partial_{\boldsymbol{\bar{p}}_{i_{1}}}
	\dots\partial_{\boldsymbol{\bar{p}}_{i_{n}}}}{m^{n}\,\imath^{n}}
	\ln \int_{\mathbb{R}^{d}}\mathrm{d}^{d}p\,e^{\imath \boldsymbol{\bar{p}}\cdot\boldsymbol{p}}
	\frac{\rho(\boldsymbol{q},\boldsymbol{p},t)}{\mu(\boldsymbol{q},t)}
	\right|_{\boldsymbol{\bar{p}}=0}
\end{equation*}        together        with        dual        tensors
$\left\{\mathsf{V}_{\boldsymbol{i}_{n}}^{\scriptscriptstyle{(n)}}\right\}_{n=0}^{\infty}$. Here
we   denote  by   $\boldsymbol{i}_{n}\equiv\,[i_{1},\dots,i_{n}]$  the
$n$-tuple  of  Euclidean indices  of  any  rank-$n$ cumulant  or  dual
tensor. If we  apply Pontryagin's principle by  looking for stationary
variations                                                          of
$\left\{\mathsf{F}_{\boldsymbol{i}_{n}}^{\scriptscriptstyle{(n)}},\mathsf{V}_{\boldsymbol{i}_{n}}^{\scriptscriptstyle{(n)}}\right\}_{n=0}^{\infty}$
and  $\mu$,  then  instead  of  the  Fokker--Planck  equation
(\ref{FP})  and  the  dynamic  programming (\ref{dp})  we  obtain  two
coupled     hierarchies    of     kinetic     equations    for     the
$\mathsf{F}^{\scriptscriptstyle{(n)}}$'s            and            the
$\mathsf{V}^{\scriptscriptstyle{(n)}}$'s and the  local equilibrium  potential $S$:
\begin{widetext}
\begin{eqnarray}
\label{BBGKY}
&&
\hspace{-0.35cm}\left(\partial_{t}+\boldsymbol{v}\cdot\partial_{\boldsymbol{q}}+\frac{n}{\tau}\right)
\mathsf{F}_{\boldsymbol{i}_{n}}^{\scriptscriptstyle{(n)}}
+ \frac{1}{m} \sum_{l=2}^{n} \binom{n}{l} \, \underset{\boldsymbol{i}_{n-l},\boldsymbol{i}_{l}}{\sym}
\mathsf{F}_{\boldsymbol{i}_{n-l},j}^{\scriptscriptstyle{(n-l+1)}}
\partial_{\boldsymbol{q}_{j}}\mathsf{F}_{\boldsymbol{i}_{l}}^{\scriptscriptstyle{(l)}}
+ \frac{1}{m} \nabla_{\boldsymbol{q}_{j}}^{\scriptscriptstyle{S}} \mathsf{F}_{j,\boldsymbol{i}_{n}}^{\scriptscriptstyle{(n+1)}}
=\delta_{n,0}\partial_{t}S
-\delta_{n,1}\partial_{\boldsymbol{q}_{i}}U
+\delta_{n,2}\frac{2\,m\,\delta_{\boldsymbol{i}_{2}}}{\beta\,\tau}
\\
&&\lefteqn{\hspace{-0.35cm}
\left(\partial_{t}+\boldsymbol{v}\cdot\partial_{\boldsymbol{q}}-\frac{n}{\tau}\right)\mathsf{V}_{\boldsymbol{i}_{n}}
^{\scriptscriptstyle{(n)}}
- \frac{1}{m} \sum_{l\geq 2} \binom{n+l-1}{l-1}
\Bigl( \frac{n}{l}\,\underset{i,\boldsymbol{i}_{n-1}}{\sym}
(\partial_{\boldsymbol{q}_{i}}
\mathsf{F}_{\boldsymbol{j}_{l}}^{\scriptscriptstyle{(l)}})\mathsf{V}_{\boldsymbol{j}_{l},\boldsymbol{i}_{n-1}}^{\scriptscriptstyle{(n+l-1)}}
- \nabla_{\boldsymbol{q}_{j}}^{\scriptscriptstyle{S}}
\mathsf{F}_{j,\boldsymbol{j}_{l-1}}^{\scriptscriptstyle{(l)}}\mathsf{V}_{\boldsymbol{j}_{l-1},\boldsymbol{i}_{n}}^{\scriptscriptstyle{(n+l-1)}}
\Bigr)
}
\nonumber\\&&
+ \frac{1}{m} \underset{i,\boldsymbol{i}_{n-1}}{\sym} 
\bigl( \partial_{\boldsymbol{q}_{i}}\mathsf{V}_{\boldsymbol{i}_{n-1}}^{\scriptscriptstyle{(n-1)}}
-n\,(\partial_{\boldsymbol{q}_{i}}\boldsymbol{v}_{j})\mathsf{V}_{j,\boldsymbol{i}_{n-1}}^{\scriptscriptstyle{(n)}} 
\bigr)
=-\frac{\delta_{\boldsymbol{i}_{2}}\,\delta_{n,2}}{\beta\,m\,\tau}
-\frac{2\,\boldsymbol{v}_{i}\,\delta_{n,1}}{\beta\,\tau}
-\delta_{n,0}
\biggl( \frac{\mathsf{F}_{j,j}^{\scriptscriptstyle{(2)}}}{m\,\tau}
+\frac{m\|\boldsymbol{v}\|^{2}}{\tau}
+\frac{2\,g\mathcal{D}_{\rho}^{\prime}}{\beta} \biggr)
\label{BBGKYd}
\end{eqnarray}
\end{widetext}  In   \eqref{BBGKY}  and  \eqref{BBGKYd}  we   use  the
conventions  that repeated  (multi-)indices  are  contracted and  that
``$\sym$'' denotes symmetrization of  the free indices in underscript.
A straightforward calculation, which  will be reported elsewhere, also
shows that \eqref{BBGKY}, \eqref{BBGKYd}  can be also derived directly
from \eqref{FP}, \eqref{dp}.  Finally,  the hierarchies are coupled by
\eqref{potential} which continues to hold with the interpretation of a
constraint               relating               the               dual
tensor~$\mathsf{V}^{\scriptscriptstyle{(1)}}$    to     the    control
potential.   \newline  The  advantage  of  introducing  \eqref{BBGKY},
\eqref{BBGKYd} is that we can use  them to approximate the solution of
our optimal  control problem  by means of  \emph{realizable closures},
\ie,  finite  order  truncations  of the  hierarchies  preserving  the
probabilistic interpretation of the cumulants \cite{Lev96}.  We notice
that truncating \eqref{BBGKY}, \eqref{BBGKYd}  at any finite order $n$
yields a controllable system of equations in the sense that the number
time derivatives in the hierarchies equals that of boundary conditions
for                               $S$                              and
$\left\{\mathsf{F}^{\scriptscriptstyle{(l)}}\right\}_{l=1}^{n}$. Furthermore,
increasing the  order of  truncation imposes  more constraints  on the
control strategy. The solution of a sequence of realizable truncations
therefore yields more and more refined lower bounds to the energy cost
\eqref{cost}.  For instance, the estimate \eqref{bound} corresponds to
the truncation of lowest order.

\section{Limit of vanishing $g$}

If we set a-priori $g=0$, by Pontryagin's principle we need to replace \eqref{potential}
with
\begin{equation*}
	U_{\star}(\boldsymbol{q},t)=\underset{U}{\operatorname{arg}\inf} \left\{
	-\int_{\mathbb{R}^{d}}\mathrm{d}^{d}q\,\mu \,\mathsf{V}^{\scriptscriptstyle{(1)}}\cdot\partial_{\boldsymbol{q}}U
	\right\}
\end{equation*}    reminiscent    of    singular    optimal    control
\cite{FlemingSoner}.   Qualitatively,  this   equation  suggests   the
decomposition  of  phase  space  into  a  ``no-action  region''  where
$\mathsf{V}^{\scriptscriptstyle{(1)}}$ vanishes and  a ``push region''
where                                                                $
\mathsf{V}^{\scriptscriptstyle{(1)}}\cdot\partial_{\boldsymbol{q}}U\,>\,0
$ and  where $\|\partial_{\boldsymbol{q}} U\|$ is  constrained only by
the  boundary  conditions.  Inspecting  \eqref{BBGKY},  \eqref{BBGKYd}
shows,      however,      that       requiring      the      condition
$\mathsf{V}^{\scriptscriptstyle{(1)}}=0$ to  hold and be  preserved by
the  dynamics  enslaves  the  macroscopic velocity  to  the  remaining
cumulants by the equation
\begin{eqnarray}
	\label{algebraic}
	&&
	\hspace{-3ex}
	\frac{2 \boldsymbol{v}_{i}}{\tau}\hspace{-0.05cm}
	=
	\hspace{-0.5ex}
	\sum_{l\geq 2} 
	\hspace{-0.5ex}
	\frac{(\partial_{\boldsymbol{q}_{i}} \mathsf{F}_{\boldsymbol{j}_{l}}^{\scriptscriptstyle{(l)}}) \mathsf{V}_{\boldsymbol{j}_{l}}^{\scriptscriptstyle{(l)}}
	- l \, \nabla_{\boldsymbol{q}_{j}}^{S}
	\mathsf{F}_{j,\boldsymbol{j}_{l-1}}^{\scriptscriptstyle{(l)}}\mathsf{V}_{\boldsymbol{j}_{l-1},i}^{\scriptscriptstyle{(l)}}
	}{m}
	\hspace{-0.05cm}-\hspace{-0.05cm}\frac{\partial_{\boldsymbol{q}_{i}}\mathsf{V}^{\scriptscriptstyle{(0)}}}{m}
\end{eqnarray}
As the boundary values of the dual tensors $\left\{\mathsf{V}_{\boldsymbol{i}_{n}}^{\scriptscriptstyle{(n)}}\right\}_{n\neq 1}$ are determined by the 
boundary conditions imposed on $\mu$ and $\left\{\mathsf{F}_{\boldsymbol{i}_{n}}^{\scriptscriptstyle{(n)}}\right\}_{n\geq 2}$, we see that in general 
(\ref{algebraic})  cannot satisfy independent boundary conditions on $\boldsymbol{v}$. 
In this sense, for $g=0$ the system is not controllable. 
The consideration of an exactly solvable case
indicates, however, that in a weaker sense $\mathsf{V}^{\scriptscriptstyle{(1)}}=0$
still governs the optimal control strategy at $g=0$.

\section{Evolution between Gaussian states} 
\begin{figure}
	\includegraphics[height=5.5cm]{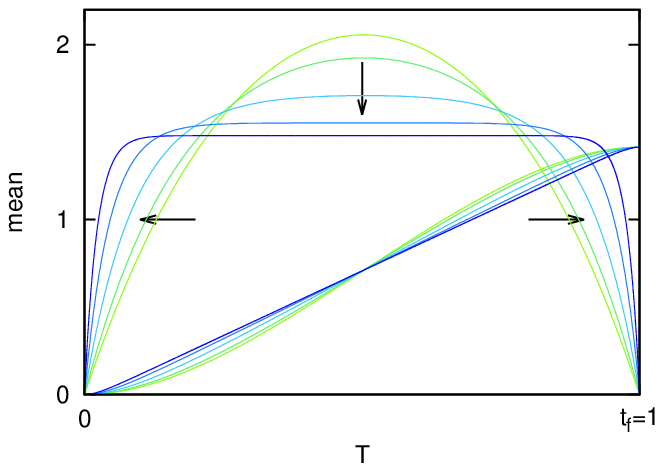}
	\caption{%
		(Color online)
		Means for Gaussian boundary conditions with parameters $\tau=1$, $m=1$, $\beta=1$, $\tf=1$, $k=1$.
		The initial and final means of $\boldsymbol{x}\equiv \left[\boldsymbol{q}\,,\boldsymbol{p}\right]$ are $(0,0)$ and $(\sqrt 2, 0)$, \resp 
		The parameter $g$ varies logarithmically from $1.28\cdot 10^{-1}$ (green/light) to $1.25 \cdot 10^{-4}$ (blue/dark). The arrows indicate the behavior for decreasing $g$. 
	}
	\label{fig:Gaussian1}
\end{figure}

\begin{figure}
	\includegraphics[height=5.5cm]{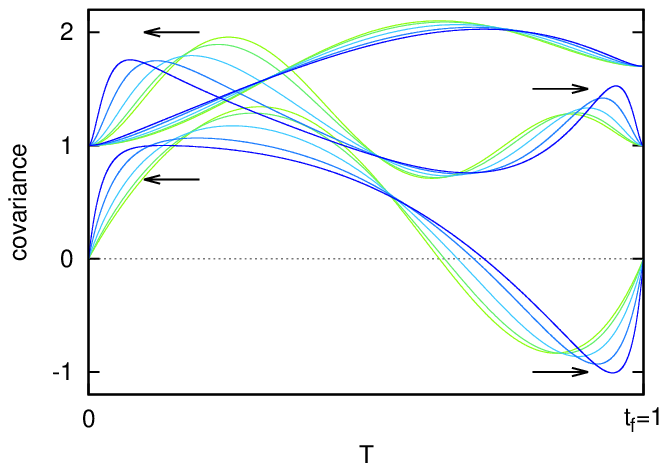}
	\caption{%
		(Color online)
		Covariances for Gaussian boundary conditions with parameters as in Figure~\ref{fig:Gaussian1}.
		The initial and final covariance matrices of $\boldsymbol{x}\equiv \left[\boldsymbol{q}\,,\boldsymbol{p}\right]$ are given by $\begin{psmallmatrix} 1 & 0 \\ 0 & 1 \end{psmallmatrix}$ and $\begin{psmallmatrix} 1.7 & 0 \\ 0 & 1 \end{psmallmatrix}$, \resp
	}
	\label{fig:Gaussian2}
\end{figure}

Let us consider transition
between  Gaussian  densities  at   finite  initial  and  final  times.
Physically this is a stylized model of a moving laser trap of changing
size (see \cite{GoScSe08} and references therein).  For these boundary
conditions  and fixed  $g>0$, the  optimal control  equations admit  a
solution in terms of a probability density which stays Gaussian in the
entire  control  horizon,  a  value  function  which  is  a  quadratic
polynomial in  $\boldsymbol{p}$ and $\boldsymbol{q}$, and  a potential
which   is  quadratic   in  $\boldsymbol{q}$.    Correspondingly,  the
hierarchies \eqref{BBGKY} and \eqref{BBGKYd} reduce to a system of $2d
\,(2d+3)$ first  order differential equations accompanied  by the same
number of  boundary conditions  for the  initial and  final cumulants.
For dimension $d=1$  the typical behavior of the solution  is shown in
Figure~\ref{fig:Gaussian1}   and  \ref{fig:Gaussian2}   for  different
values of $g$.

As $g \to 0$ the limit behavior of the cumulants is described by the ``slow manifold'' specified by
    the condition $\mathsf{V}^{\scriptscriptstyle{(1)}}=0$ and the evolution law (\ref{algebraic}). 
    Only in a layer close to the boundaries of the control horizon 
    cumulants get away from the slow manifold along exponentially stable and unstable directions with
    rates of the order $O(1/\sqrt{g})$ in order to satisfy the boundary conditions. 
    The description of singular boundary value problems in terms of invariant manifolds is well known 
    in the theory of dynamical systems \cite{TiKoJo94}.
    We refer the interested reader to \cite{PMGSc14b} for the details of the multiscale expansion
    \cite{PaSt08} proving the foregoing qualitative picture.

    \section{Overdamped limit} 

The relevance  of the ``slow manifold''
    condition $\mathsf{V}^{\scriptscriptstyle(1)}=0$  appears from the
    fact  that   it  permits   to  recover   directly  at   $g=0$  the
    ``overdamped''  limit  of  the  Langevin--Kramers  dynamics.   The
    overdamped regime  corresponds to the  assumption of a  wide scale
    separation between  the control  horizon $[0,\tf]$ and  the Stokes
    time $\tau$ and between the characteristic length scale $L$ of the
    configuration         space         boundary        data         $
    \mu_{\mathrm{\iota}}(\boldsymbol{q}/L)$,
    $\mu_{\mathrm{f}}(\boldsymbol{q}/L)$   and  the   typical
    length  scale  $\ell=\tau/\sqrt{\beta\,m  }$ of  the  uncontrolled
    process. In  the overdamped  regime, the  quantifier of  the scale
    separation          is          the         Stokes          number
    $\st=\tau/\tf=\ell^{2}/L^{2}\ll\,1$.   Under these  hypotheses, we
    can look for  an asymptotic solution of  (\ref{FP}) and (\ref{dp})
    by  expanding  around  a  Maxwell--Boltzmann  momentum  equilibrium
    distribution       $\mu_{\scriptscriptstyle{MB}}(\boldsymbol{p})$
    perturbed              at               large              scales,
    $(\boldsymbol{\tilde{q}},\tilde{t})\equiv(\sqrt{
    \st}\,\boldsymbol{q},\st t)$, by the action of a control potential
    of     the     form    $U(\boldsymbol{q},t)\equiv     U_{0}(\sqrt{
    \st}\,\boldsymbol{q},\st   t)+O\left(\sqrt{  \st}\right)$.  From now on we specify by an 
    underscript the order of the perturbative expansion. 
    Upon setting
    $\partial_{\boldsymbol{\tilde{q}}}S_{0}=-\partial_{\boldsymbol{\tilde{q}}}\ln
    \mu_{0}$,  and applying  standard homogenization  techniques (see
    e.g. \cite{PaSt08}, see also \cite{PMG14}) we get for the solution
    of the Fokker--Planck equation \eqref{FP}:
    \begin{eqnarray}
            \label{centeringFP}
            &&\rho(\boldsymbol{p},\boldsymbol{\tilde{q}},\tilde{t})=
            \mu_{\scriptscriptstyle{MB}}(\boldsymbol{p})\,\mu_{0}(\boldsymbol{\tilde{q}},\tilde{t})\times
            \nonumber\\&&
            \Bigl(
            1 + \sqrt{ \st}\,
	\tfrac{\tau}{m} \,\boldsymbol{p} \cdot \partial_{\boldsymbol{\tilde{q}}} 
         (S_{0} - \beta\,U_{0} )(\boldsymbol{\tilde{q}},\tilde{t})
	+O(\st) \Bigr)
\end{eqnarray}
For the solution of the dynamic programming \eqref{dp} we obtain
\begin{eqnarray}
	\label{centeringV}
	&&V(\boldsymbol{p},\boldsymbol{\tilde{q}},t,\tilde{t})
	=
	\frac{(\tf-t)\,d}{\beta\,\tau} + \frac{\|\boldsymbol{p}\|^{2}}{2\,m}
	+ V_{0}(\boldsymbol{\tilde{q}},\tilde{t}) +
	\nonumber\\&&
	\sqrt{\st} \Bigl( V_{1}(\boldsymbol{\tilde{q}},\tilde{t})
	+ \tfrac{\tau}{m} \,\boldsymbol{p} \cdot 
        \partial_{\boldsymbol{\tilde{q}}}\,(V_{0}-U_{0})(\boldsymbol{\tilde{q}},\tilde{t}) \Bigr)
	+O(\st)
\end{eqnarray}  The functions  $S_{0}$ in  \eqref{centeringFP}
and  $V_{0}$  in  \eqref{centeringV},  respectively,  obey  the  local
equilibrium potential equation
\begin{equation*}
	\partial_{\tilde{t}}S_{0}-\tfrac{\tau}{m}       \Bigl(
(\partial_{\boldsymbol{\tilde{q}}}S_{0})
\cdot\partial_{\boldsymbol{\tilde{q}}}
-    \partial_{\boldsymbol{\tilde{q}}}^{2}    \Bigr)   (    U_{0}    -
\tfrac{1}{\beta}S_{0} ) = 0
\end{equation*}
and the dynamic programming equation
\begin{equation*}
	\partial_{\tilde{t}}V_{0}-\tfrac{\tau}{m} 
	\Bigl( (\partial_{\boldsymbol{\tilde{q}}}U_{0}) \cdot \partial_{\boldsymbol{\tilde{q}}}
	- \tfrac{1}{\beta} \partial_{\boldsymbol{\tilde{q}}}^{2} \Bigr) (V_{0}-U_{0})=0
\end{equation*} These  equations specifying  two of the  three optimal
control   equations  governing   the   minimal  heat   release  by   a
Langevin--Smoluchowski  dynamics   between  $\mu_{\iota}$  and
$\mu_{\mathrm{f}}$ \cite{AuMeMG11}.  In  order to recover the
third condition, we use  \eqref{centeringFP} and \eqref{centeringV} to
evaluate
\begin{equation*}
	\mathsf{V}^{\scriptscriptstyle{(1)}}(\boldsymbol{\tilde{q}},\tilde{t})
	=
	-\sqrt{\st} \, \tfrac{\tau}{m} \, \partial_{\boldsymbol{\tilde{q}}} \Bigl( 2 \, U_{0} - \frac{S_{0}}{\beta}
	- V_{0} \Bigr)
	(\boldsymbol{\tilde{q}},\tilde{t})+O(\varepsilon)
	\nonumber
\end{equation*}             Then             the             condition
$\mathsf{V}^{\scriptscriptstyle{(1)}}=0$  yields exactly  the relation
between $U_{0}$, $V_{0}$,  $S_{0}$ that allows us to  recover the very
same Monge--Amp\`ere--Kantorovich equations of \cite{AuMeMG11}:
\begin{gather*}
	\partial_{\tilde{t}}\tilde{U}
	-\tfrac{\tau}{2\,m}\partial_{\boldsymbol{\tilde{q}}}\tilde{U}\cdot\partial_{\boldsymbol{\tilde{q}}}\,\tilde{U}
	\\
	\partial_{\tilde{t}}S_{0} - \tfrac{\tau}{m} (\partial_{\boldsymbol{\tilde{q}}}S_{0})\cdot\partial_{\boldsymbol{\tilde{q}}}
	\tilde{U} + \tfrac{\tau}{m} \partial_{\boldsymbol{\tilde{q}}}^{2}
	\tilde{U} = 0
\end{gather*}
with $\tilde{U}\equiv U_{0}-S_{0}/\beta$.

\section{Conclusion}   

We showed  how Pontryagin's  principle can be  used to  derive refined
bounds  for  the   Second  Law  of  thermodynamics  in   the  case  of
nanomechanical  systems.   We  also  established  a  relation  between
optimal control and kinetic theory,  which renders available ideas and
tools of dilute gas \cite{CeIlPu94,Lev96} and optimal transport theory
\cite{Villani} to  the construction of optimal  protocols implementing
at  the  nano-scale  information  processing operations  such  as  the
erasure of a bit \cite{DiLu09}.

The  authors  acknowledge the  Finnish  Academy  Center of  Excellence
``\emph{Analysis and Dynamics}'' for support. The authors are grateful
to  Carlos  Mej\'{i}a-Monasterio  and   Antti  Kupiainen  for  helpful
comments.

\addcontentsline{toc}{section}{Bibliography}
\bibliography{/home/paolo/RESEARCH/BIBTEX/jabref}{} 
\bibliographystyle{aipnum4-1} 
\end{document}